\documentclass[twocolumn,10pt]{article}
\usepackage[a4paper,margin=0.7in]{geometry}
\usepackage{float}
\usepackage[dvipsnames]{xcolor}
\usepackage{graphicx}

\usepackage{authblk}  

\usepackage{titlesec} 
\usepackage{abstract} 
\usepackage[numbers,sort&compress]{natbib} 
\usepackage{amsmath, amsthm, amssymb}
\usepackage{romannum}
\usepackage{amsfonts}
\usepackage{tikz}

\renewcommand{\thesection}{\Roman{section}}
\renewcommand{\thesubsection}{\Roman{section}.\Alph{subsection}}
\titleformat{\section}{\large\bfseries\fontsize{10}{13}\selectfont\MakeUppercase}{\thesection.}{0.5em}{}
\titleformat{\subsection}{\normalsize\bfseries}{\thesubsection.}{0.5em}{}
\usepackage{caption}
\usepackage{subcaption}
\usepackage{hyperref}
\hypersetup{breaklinks=true}
\hypersetup{
    colorlinks=true,
    linkcolor=blue,
    filecolor=blue,      
    urlcolor=blue,
    citecolor=blue,
    linkcolor=red,
}

\title{\textbf{Quantum Fourier Transform Infrared Spectroscopy: Evaluation, Benchmarking and Prospects}}
\author[1]{Paul Gattinger\thanks{Corresponding Author: paul.gattinger@recendt.at}}
\author[2]{Andreas W. Schell}
\author[3]{Sven Ramelow}
\author[1]{Markus Brandstetter}
\author[1]{Ivan Zorin}

\affil[1]{Research Center for Non-Destructive Testing, Science Park 2, Altenberger Str. 69, 4040 Linz, Austria}
\affil[2]{Institute of Semiconductor and Solid State Physics, Johannes Kepler University, Altenberger Str. 69, 4040 Linz, Austria}
\affil[3]{Institut f{\"u}r Physik, Humboldt-Universit{\"a}t zu Berlin, Newtonstr. 15, 12489 Berlin, Germany}
\date{\vspace{-1em}\fontsize{9}{12}\selectfont\today\vspace{-2em}}

\begin{document}
\twocolumn[
\maketitle

\begin{onecolabstract}  
Sensing with undetected photons has enabled new, unconventional approaches to Fourier transform infrared (FTIR) spectroscopy. Leveraging properties of non-degenerated entangled photon pairs, mid-IR information can be accessed in the near-IR spectral domain to perform mid-IR spectroscopy with silicon-based detection schemes. 
Here, we address practical aspects of vibrational spectroscopy with undetected photons using a quantum-FTIR (QFTIR) implementation. The system operates in the spectral range from around 3000~cm\textsuperscript{-1} to 2380~cm\textsuperscript{-1} (detection at around 12500~cm\textsuperscript{-1}) and possesses only 68~pW of mid-IR probing power for spectroscopic measurements with a power-dependence of the signal-to-noise ratio of 1.5$\cdot$10\textsuperscript{5} mW\textsuperscript{-1/2}. We evaluate the system's short- and long-term stability and experimentally compare it to a commercial FTIR instrument using Allan-Werle plots to benchmark our QFTIR implementation's overall performance and stability. In addition, comparative qualitative spectroscopic measurements of polymer thin films are performed using the QFTIR spectrometer and a commercial FTIR with identical resolution and integration times. Our results show under which conditions QFTIR can practically be competitive or potentially outperform conventional FTIR technology.
\end{onecolabstract}
\vspace{0.5em}
]\saythanks

\section{\label{sec:Intro}Introduction}
Fourier transform infrared (FTIR) spectroscopy is a well-established, non-destructive analytical method that has been a gold standard in vibrational spectroscopy for over half a century~\cite{griffiths2007fourier,chalmers_handbook_2001}. Typical tabletop FTIR spectrometers consist of a light source, a detector, and an interferometer (Michelson or Mach-Zehnder type). Thermal light sources are still widespread in standard instruments due to their broadband coverage, low costs, and stability. However, emerging laser sources are a promising enhancement for FTIR spectroscopy due to their spatially coherent emission and high optical brightness~\cite{Zorin:22,Krebbers:24}. In terms of detection, standard low band-gap infrared (IR) detectors usually suffer from several types of noise (e.g., thermal, $1/f$) and, therefore, low detectivities~\cite{Rogalski2010}. Even well-developed standard mercury cadmium telluride (HgCdTe, MCT) detectors, which usually have to be cryogenically cooled, still have three to four orders of magnitude higher noise equivalent powers than Si-based detector schemes~\cite{Rogalski2021,Konstantatos2018}. Thus, further refinement of these classical methods is technically challenging because it faces fundamental limitations of the IR source and detector technologies.

The recently emerging field of metrology with undetected photons has the potential to circumvent the need for classical IR sources and detectors in IR spectroscopy. The fundamental metrology principles rely on nonlinear interferometry~\cite{Chekhova2016} and exploit induced coherence of quantum sources without induced emission~\cite{Zou1991,Mandel1991}. Initial publications on the use of these methods in the near-IR demonstrated the possibility of transferring information between correlated, non-degenerate photon pairs created via spontaneous parametric down-conversion (SPDC)~\cite{Lemos2014,BarretoLemos2022}. However, the underlying concept can also be applied to more widely separated non-degenerate entangled photon pairs. By using properties of specifically engineered broadband SPDC light sources, mid-IR photons can be used to probe samples, while detection is carried out by measuring their entangled partner photons in the near-IR spectral domain. This allows experimentalists to utilize Si-based detector technology to effectively measure in the mid-IR regime, normally only accessible by mid-IR detectors (InSb, HgCdTe, PbSe, etc.). After initial demonstrations of IR imaging, the methods have been adapted for spectroscopy~\cite{Kalashnikov2016,Paterova2022,Kaufmann2022,Cardoso2024}, optical coherence tomography\cite{Valles2018,Paterova2018,Vanselow2020} and, particularly, for quantum FTIR (QFTIR) spectroscopy~\cite{Lindner2020,Lindner:21,Lindner2022,Mukai2022,Tashima2024}, yielding cost-effective and promising alternative solutions for infrared spectroscopy.

In this manuscript, we evaluate the applied spectroscopic performance of novel QFTIR solutions, including aspects of noise and stability. We benchmark an experimental QFTIR system against a state-of-the-art FTIR spectrometer and emphasize the distinct features of these non-classical methods. Based on the results, we identify potential application scenarios and prospects of QFTIR spectroscopy with undetected photons. The bi-photons in our QFTIR spectrometer are generated in a 2.55~mm periodically poled potassium titanyl phosphate crystal (ppKTP). The sensing range of our system is from 780~nm to around 820~nm, and the photons used to probe absorption span the range from around 3000~cm\textsuperscript{-1} to 2380~cm\textsuperscript{-1}, with only 68~pW of mid-IR probing power. The implemented system possesses and introduces original engineering solutions, such as scanning of the combined pump and signal phases and a reference interferometric system, and is aligned and adapted to meet the interests of IR spectroscopy.

\section{\label{sec:nli}Nonlinear interferometry and sensing with undetected photons}

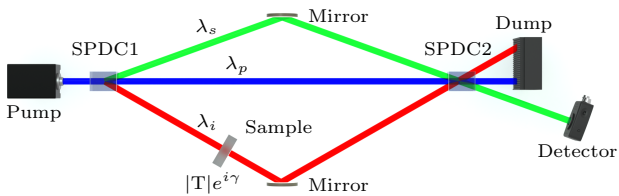
\begin{figure}[ht]
\centering
\begin{tikzpicture}
  \node[anchor=south west,inner sep=0] (image) at (0,0,0) {\includegraphics[width=0.45\textwidth]{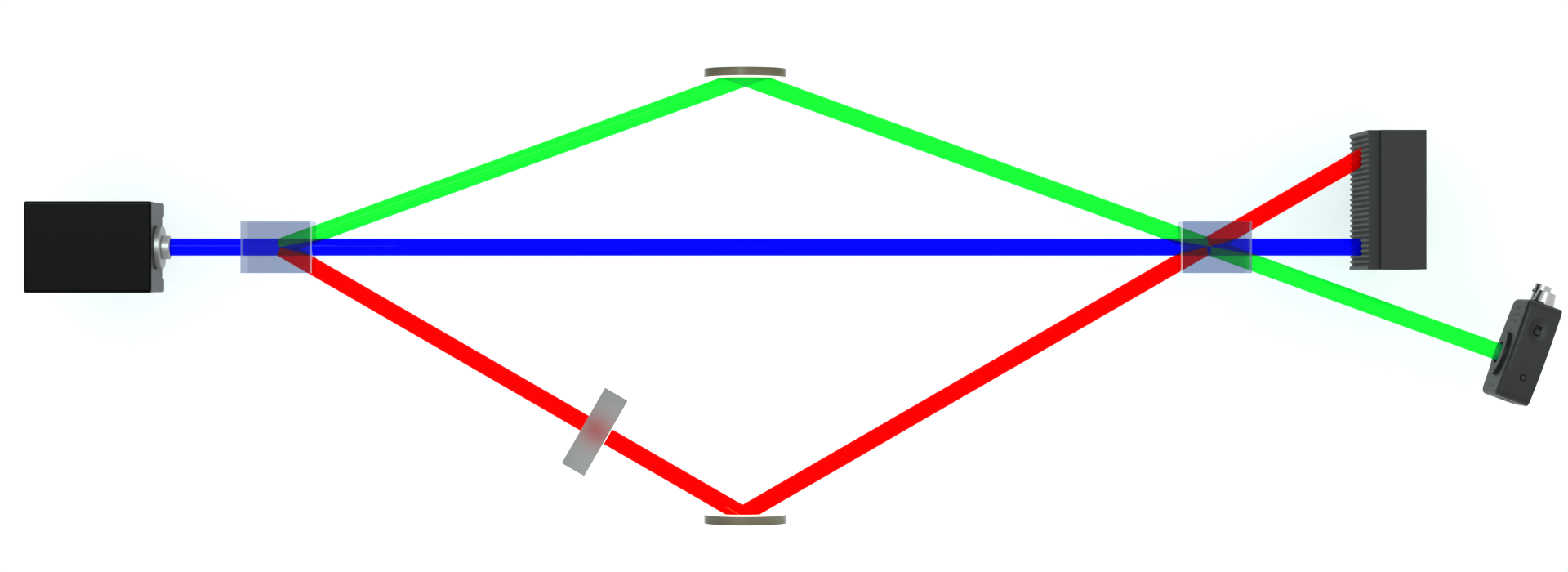}};
  \begin{scope}[x={(image.south east)},y={(image.north west)}]
    \draw (0.06,0.42) node[]{\color{black}\scriptsize Pump};
    \draw (0.18,0.72) node[]{\color{black}\scriptsize SPDC1};
    \draw (0.77,0.72) node[]{\color{black}\scriptsize SPDC2};
    \draw (0.57,0.87) node[]{\color{black}\scriptsize Mirror};
    \draw (0.57,0.09) node[]{\color{black}\scriptsize Mirror};
    \draw (0.97,0.25) node[]{\color{black}\scriptsize Detector};
    \draw (0.47,0.35) node[]{\color{black}\scriptsize Sample};
    \draw (0.88,0.83) node[]{\color{black}\scriptsize Dump};
    \draw (0.4,0.64) node[]{\color{black}\scriptsize $\lambda_p$};
    \draw (0.35,0.82) node[]{\color{black}\scriptsize $\lambda_s$};
    \draw (0.35,0.38) node[]{\color{black}\scriptsize $\lambda_i$};
    \draw (0.36,0.1) node[]{\color{black}\scriptsize $|\mathrm{T}|e^{i\gamma}$};
  \end{scope}
\end{tikzpicture}
\caption{\label{fig:nl_interferometry} Principles of nonlinear interferometry: a common pump ($\lambda_p$) illuminates two nonlinear crystals triggering generation of correlated (entangled) photon pairs ($\lambda_s$ and $\lambda_i$) via spontaneous parametric down-conversion (SPDC) in each crystal; information on which photon comes from which source is erased after the second crystal (SPDC2) due to the indistinguishability of the photon paths allowing for interference between the probability amplitudes of the photons coming from the first or second SPDC source; the resulting interference pattern when scanning the path length difference (akin to conventional interference) is identically present for both signal and idler and can thus be detected by only measuring in the $\lambda_s$ signal domain, i.e., mid-IR information can be detected in the near-IR without exposing the sample to near-IR light.}
\end{figure}

In contrast to classical FTIR spectroscopy exploiting classical optical interferometry, QFTIR spectroscopy relies on the principles of nonlinear interferometry and sensing with undetected photons. A basic nonlinear interferometer is schematically depicted in Figure~\ref{fig:nl_interferometry}. A monochromatic laser at wavelength $\lambda_p$ pumps two crystals possessing a $\chi^{(2)}$ nonlinearity. The quantum mechanical process of SPDC enables the generation of so-called signal ($\lambda_s$) and idler ($\lambda_i$) photons in the first crystal (SPDC1), which are, in general, correlated in energy and momentum. 
The signal and idler beams propagate under different angles (in the case of critical phasematching); a sample with transmissivity $|T|$ is inserted in the idler beam path. Subsequently, both $\lambda_s$ and $\lambda_i$ photons are reflected by mirrors and overlapped with the pump beam in the second crystal (SPDC2) where, again, $\lambda_s$ and $\lambda_i$ photons could be created. The idler photons are discarded, and the signal photons are detected. If the $\lambda_i$ and $\lambda_s$ photon paths are indistinguishable, i.e. if the nonlinear interferometer is well aligned, coherence is induced~\cite{Zou1991,Mandel1991, Chekhova2016} and thus interference occurs. Assuming no losses for the pump and signal beams the normalized photon rate at the detector then follows~\cite{BarretoLemos2022}: 
\begin{equation}\label{eqs:interference}
R_{s} = \frac{1 + |T| \cos(\Delta \phi + \gamma)}{2},
\end{equation}
where $|T|$ is the absolute value of the transmission through the sample, $\Delta \phi$ is the phase shift between pump, signal, and idler and $\gamma$ is the phase delay introduced by the sample, and thus allows phase-sensing. Transmission sensing can be exemplified with a beam block ($|T| = 0$) inserted in the idler path, for which the interference of the signal photons would disappear. The signal photons created in the first SPDC crystal would, in principle, be distinguishable from the signal photons created in the second SPDC crystal as a detector or the beam block could be used to make this distinction. This theoretical distinguishability leads to the vanishing of the interference. This holds true for a partial beam-block resulting in partial distinguishabily and thus partial interference. Thus, the interference contrast in the $\lambda_s$ region reflects the absorptive and dispersive properties given in the correlated spectral region of undetected photons $\lambda_i$.

Many different nonlinear crystals can be used to generate SPDC. However, in the following, we will focus on periodically poled crystals. These crystals have periodically alternating crystal domains, which leads to effectively phase-matched and thus efficient quasi-collinear emission of pump, signal, and idler photons (quasi-phase matching). 

For the generated photon pairs, energy conservation has to be fulfilled:
\begin{equation} \label{eq:energy_cons}
\omega_p = \omega_s + \omega_i,
\end{equation}
where $\omega_{p,s,i}$ represent the angular frequencies of pump, signal, and idler photons, as well as the phase-matching condition, which accounts for momentum conservation and is given by:
\begin{equation}\label{eq:QPM}
\mathbf{\Delta k} = \mathbf{k_p} - \mathbf{k_s} - \mathbf{k_i} - \mathbf{k_\Lambda}, 
\end{equation}
where $\mathbf{\Delta k}$ is the phase mismatch that is subject to minimization, $\mathbf{k_{p,s,i}}$ are the corresponding wavevectors, and $\mathbf{k_\Lambda}$ is the effective crystal vector contribution due to the periodic poling. Along the axis of the crystal $\mathbf{k_\Lambda}$ takes a scalar form of $2\pi/\Lambda$, where $\Lambda$ is the poling period. 
A coherent buildup of signal and idler fields can only be achieved if their phases are in sync with each other. Tuning of the quasi-phase matching condition to match group velocities of the signal and the idler fields\cite{Vanselow2019} allows spectrally widely separated broadband signal (near-IR) and idler (mid-IR) photons to be generated. This broadband emission enables broad-band spectroscopy using FT techniques: 
An interferogram with a certain visibility (akin to classical FT broadband interferograms) can be observed by scanning the pathlength-difference of the signal or idler beams. The magnitude of the visibility is proportional to the degree of (partial) indistinguishability of the signal and idler photon paths~\cite{Mandel1991}. The interference will (partially) vanish if the idler photons in the idler arm are (partially) absorbed. 
The emergent interference is determined by the relative phase of all three fields involved, which is defined as follows:
\begin{equation} \label{eq:phase}
\Delta\phi = \phi_p-\phi_s-\phi_i,
\end{equation}
where $\phi_{p,s,i}$ are the phases of pump, signal, and idler photons respectively. If the scanned path contains both the signal and the pump, the difference of $\phi_s$ and $\phi_p$ is scanned. Contrary to intuitive belief about the instability of scanning short-wavelength interferometric arms, scanning both fields and their relative phase difference makes this configuration quantitatively equivalent to scanning the mid-IR idler arm length. This facilitates alignment as well as potential adaptation of the system for QFTIR imaging \cite{Placke2023}.

\section{\label{sec:exp}Experimental realization}

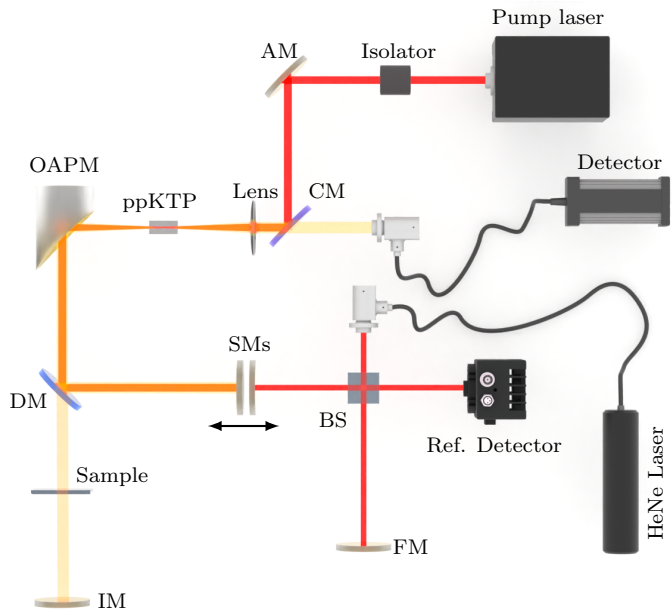
\begin{figure*}[ht]
\centering
\begin{tikzpicture}
  \node[anchor=south west,inner sep=0] (image) at (0,0,0) {\includegraphics[width=0.9\textwidth]{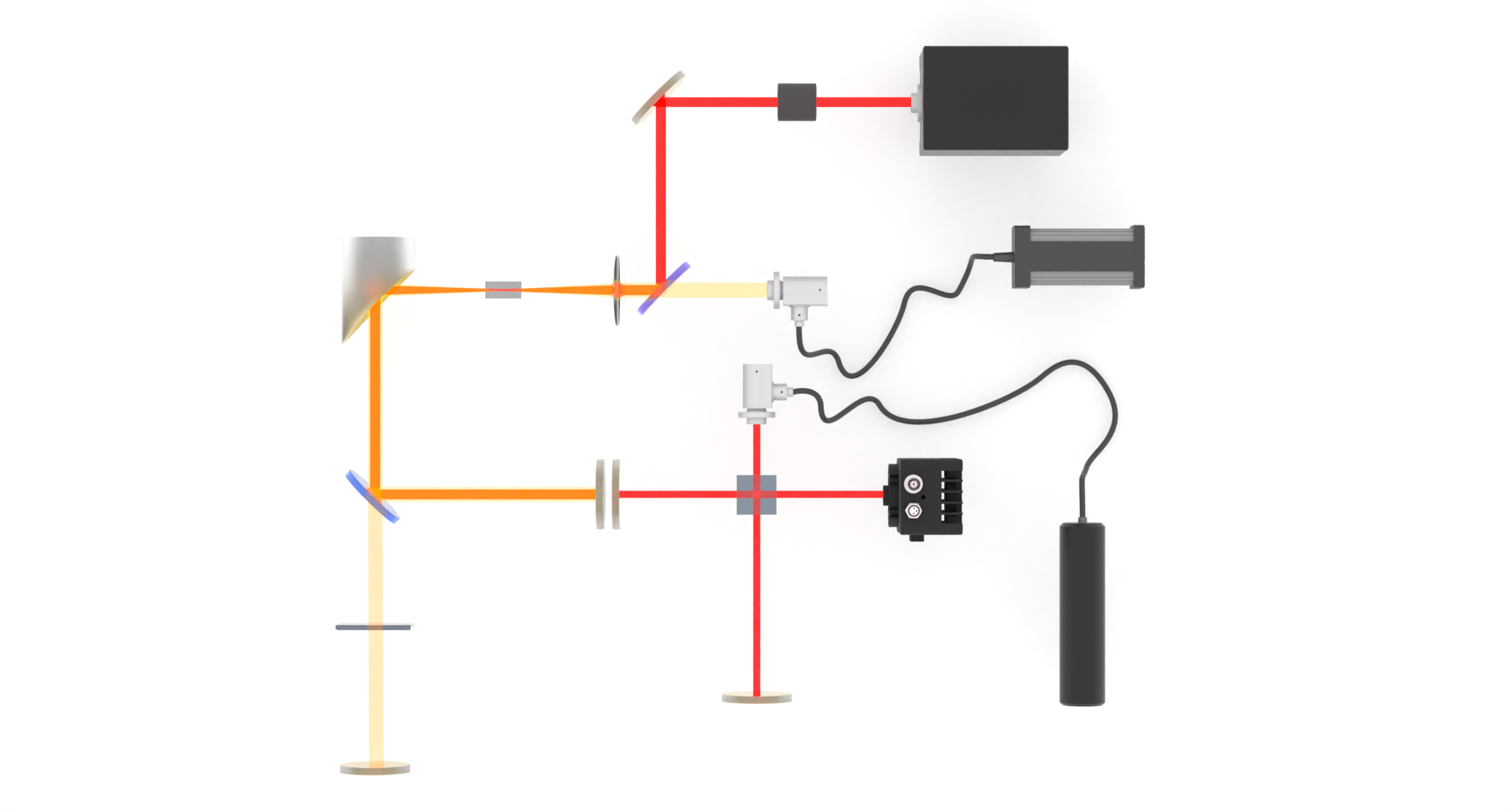}};
  \begin{scope}[x={(image.south east)},y={(image.north west)}]
    \draw (0.655,0.97) node[]{\color{black}\footnotesize Pump laser};
    \draw (0.715,0.745) node[]{\color{black}\footnotesize Detector};
    \draw (0.53,0.92) node[]{\color{black}\footnotesize Isolator};
    \draw (0.43,0.92) node[]{\color{black}\footnotesize AM};
    \draw (0.33,0.68) node[]{\color{black}\footnotesize ppKTP};
    \draw (0.25,0.74) node[]{\color{black}\footnotesize OAPM};
    \draw (0.41,0.7) node[]{\color{black}\footnotesize Lens};
    \draw (0.47,0.7) node[]{\color{black}\footnotesize CM};
    \draw (0.22,0.37) node[]{\color{black}\footnotesize DM};
    \draw (0.405,0.46) node[]{\color{black}\footnotesize SMs};
    \draw (0.29,0.055) node[]{\color{black}\footnotesize IM};
    \draw (0.475,0.34) node[]{\color{black}\footnotesize BS};
    \draw (0.54,0.14) node[]{\color{black}\footnotesize FM};
    \draw (0.29,0.25) node[]{\color{black}\footnotesize Sample};
    \draw (0.61,0.3) node[]{\color{black}\footnotesize Ref. Detector};
    \draw (0.745,0.24) node[rotate=90]{\color{black}\footnotesize HeNe Laser};
    \draw [latex-latex,thick, color={black}] (0.40-0.03, 0.33) -- (0.40+0.03, 0.33);
  \end{scope}
\end{tikzpicture}
\caption{\label{fig:setup} Experimental scheme of the quantum Fourier transform infrared spectrometer; the nonlinear interferometer comprises: 500 mW pump laser (660~nm), optical isolator, AM - alignment mirror, CM - cold mirror, Lens - achromatic lens, ppKTP - periodically poled potassium titanyl phosphate, OPM - off-axis parabolic mirror, DM - dichroic mirror, IM - idler (mid-IR) arm mirror, SM - scanning mirror; the scanning is controlled and linearized using a reference classical monochromatic interferometer formed by: a HeNe laser (632.8~nm), BS - beam splitter, FM - fixed mirror, and reference detector; the scanning mirrors of both interferometers are coupled and scanned simultaneously.}
\end{figure*}

Similar to the standard classical FTIR spectrometer, which is based on the Michelson interferometer, a QFTIR spectrometer can be implemented in a Michelson-type nonlinear interferometer configuration\cite{Lindner2022, Mukai2022}. It is essentially a folded nonlinear interferometer shown in Fig.~\ref{fig:nl_interferometry}, where the pump photons pass through the same crystal twice. 
Instead of a classical amplitude beam splitter, a dichroic mirror is employed to separate the signal and pump fields from the idler field. The signal and pump photons propagate in the signal arm of the interferometer while the idler photons propagate in the idler arm. 
All three fields are recombined at the dichroic mirror and pass through the nonlinear crystal a second time, afterwards, only the spectrally separated signal photons are detected. 

A scheme of the demonstrator setup is shown in Figure \ref{fig:setup}. A single mode (spatial and longitudinal) visible, linearly polarized laser (660~nm, 500~mW, Cobolt Flamenco) serves as pump laser. To avoid back reflections into the laser cavity, an optical isolator based on a Faraday rotator is employed. The pump beam is then guided into the nonlinear interferometer via a dichroic mirror (cold mirror CM, Thorlabs M254C45), which also separates the pump path and the path for signal detection later on. The pump beam is expanded by a telescope to 2.3~mm (not shown in the figure). A 75~mm achromatic lens (Thorlabs AC254-075-B) is then used to focus the pump beam into a periodically poled potassium titanyl phosphate (ppKTP) crystal ($\Lambda = 20.45$~\textmu m, $l=2.55$~mm, Raicol) whose domain orientations are set to match the polarization of the laser. For the given optics and crystal length the optimized focusing parameter is 1.42 (ratio of the crystal length and the confocal parameter). Broadband signal and idler photons are created via SPDC. Together with the pump beam, the emerging signal and idler beams are collimated by an off-axis parabolic mirror (OAPM, $\text{RFL}=76.1$~mm, Thorlabs MPD139-P01) to avoid chromatic aberrations. A custom dichroic mirror (DM, Photon Laseroptik) separates the pump and signal (reflected) from the idler photons (transmitted), resulting in the formation of combined pump and signal and idler interferometer arms, respectively. The mirror in the combined pump and signal arm (SM) is mounted on an automated linear stage (Physik Instrumente, M-126.PD1), while the mirror that closes the idler arm (IM) is fixed. For transmission measurements, the sample may be placed anywhere in the idler arm. All the reflected fields are recombined at the dichroic mirror and focussed again into the ppKTP crystal, where signal and idler photons are created in a second SPDC process. 
If the nonlinear interferometer is well aligned, i.e., a high degree of transverse mode overlap is maintained, it, in principle, cannot be distinguished, whether the detected signal photons were created in the first pass through the nonlinear crystal or in the second pass. This means interference can occur according to Eq.~\ref{eqs:interference}. 
After the second pass through the crystal, the signal photons are separated from the pump and transmitted through the cold mirror towards the detection arm; the optical isolator blocks the pump reflected by the cold mirror. The idler photons are discarded (absorbed, scattered) at the cold mirror, however, the spectroscopic information is also present in the signal spectral range. A notch filter (Thorlabs NF658-26) is placed after the cold mirror to block residual pump radiation. The signal photons carrying information about the phase-shift and absorption from the sample in the mid-IR range are coupled into a single-mode fiber via an achromatic doublet ($f=19$~mm, Thorlabs, AC127-019-B-ML), and a Photodetector (Femto OE-200-SI-FC) is used for detection. The interferograms are recorded with an oscilloscope (Teledyne Lecroy, HDO6104A). A reference interferometer based on a HeNe laser is attached to the back side of the scanning mirror to track its movement for postprocessing. Linearization and resampling of the interferograms, as well as apodization (Blackman Harris), zero filling, and Fourier transformation are carried out in a Python-based software. Consequently, the room-temperature QFTIR system provides classical broadband mid-IR spectra with no mid-IR optics, sources, or detectors in its design.

\section{\label{sec:res}Results}

In order to assess the system's performance and draw comparisons to standard tabletop instruments, several comparative measurements were made. All QFTIR and FTIR measurements displayed in this section were executed at 4~cm$^{-1}$ resolution and a single spectrum acquisition time of 1~s. 

\begin{figure}[ht]
\centering
\includegraphics[width=0.99\linewidth]{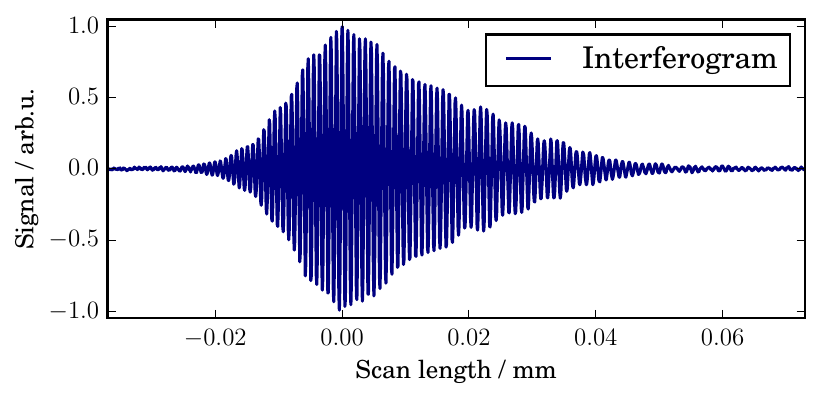}
         \caption{Typical QFTIR interferogram, transmission through air, baseline-corrected; the asymmetric shape is due to an uncompensated dispersion imbalance between the interferometric arms due to differences in the spectral regions of the probing and sensing fields.}
         \label{fig:interferogram}
\end{figure}

A typical interferogram containing the normalized QFTIR central burst, used to reconstruct mid-IR spectra is displayed in Figure~\ref{fig:interferogram}. 

\begin{figure}[ht]
\centering
\includegraphics[width=0.99\linewidth]{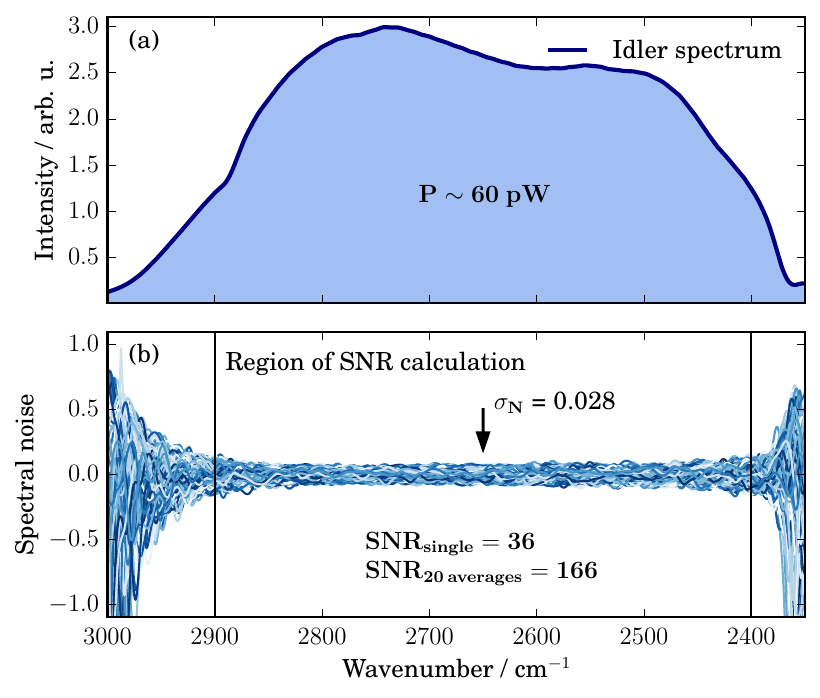}
         \caption{\label{fig:noise} Spectroscopic short-term performance a): Idler spectrum of the crystal employed for measurements; the power of the sensed idler emission is around 60~pW; absorption of ambient CO$_2$ limits the spectral bandwidth at 2350~cm$^{-1}$; b): Assessment of the shot-to-shot spectral noise calculated using subsequently measured idler spectra; the SNR is calculated for the derived standard deviation ($\sigma_N$) of the noise floor.}
         \label{fig:spec_noise}
\end{figure}

Nonlinear interferometers are unbalanced, i.e., group velocity dispersion that is introduced between signal and idler fields by the crystal, and optics is not naturally compensated. Therefore, the interferograms are generally (if no optical correction is made) asymmetric. The single channel transmission spectrum, as shown in Figure~\ref{fig:spec_noise}(a), is reconstructed using standard FT spectroscopy reconstruction algorithms; the phase spectrum is omitted here, but can also yield spectroscopic information\cite{Lindner2022}.

It should be noted that the probability of the SPDC process\textemdash normalized to the number of pump photons is typically very low and at the level of ~10\textsuperscript{-9} in the present setup, i.e., the generated signal and pump fields are extremely weak. Therefore, the power in the mid-IR arm can not be measured directly using classical detectors. The sensing power in the mid-IR range was determined by measuring the free-space power of the signal beam using a high-sensitivity Si detector (quantitative calibrated responsitivity $10^{11}$~V/W) to be 645~pW. This corresponds to a photon-pair-rate of $2.59\times10^{9}$~s$^-1$. The integrated beam power of the mid-IR beam probing the sample was then inferred to be around 68~pW (half of the total mid-IR power) using energy conservation, i.e., directly relying on the quantum principles that the near-IR and mid-IR photons are strictly generated only in pairs. These operating power levels emphasize a unique feature of QFTIR\textemdash the ability to operate at extremely low levels, which is of interest for a variety of spectroscopic applications. However, it imposes certain limitations on the fundamental signal-to-noise (SNR) ratio constraints, which in this regime is dominated by the fundamental shot-noise of the light itself. In the QFTIR setting, the signal photons were coupled into a single-mode fiber to minimize stray light. The coupling efficiency was 0.875, resulting in around 565~pW of sensed near-IR light and, therefore, effectively 60~pW of mid-IR power.

To evaluate the SNR of the QFTIR system and compare it to classical instruments, well-adopted FTIR standards and methods were employed\cite{Saptari2003,griffiths2007fourier}. A set of single-channel spectra with no sample and no changes to the optical path was recorded. The spectral noise $N$ was calculated according to $N = 1 - I_1/I_2$, where $I_{1,2}$ are two subsequently recorded single channel transmission spectra. Figure~\ref{fig:spec_noise} shows the SPDC idler spectrum and illustrates the derivation of the measured SNR ratio. In Fig.~\ref{fig:spec_noise}(b) the spectral noise curves that were recorded with the QFTIR system are shown. The SNR was calculated as $1 / \sigma_N$, where $\sigma_N$ is the standard deviation of the spectral noise in the range between 2900~cm$^{-1}$ and 2400~cm$^{-1}$. The spectral range used for the calculation is reasonably defined based on the significance of the SPDC idler spectrum (shown in Fig.~\ref{fig:spec_noise}(a)). For single-shot transmission spectra, the SNR was calculated to be 36, whereas for twenty averages a value of 166 could be reached, which agrees very well with the expected square-root scaling within the statistical error. The theoretical shot-noise limited performance of the instrument for the given parameters can be estimated using a formula reported by Lindner et al.\cite{Lindner:21}. The provided framework accounts for photon shot noise for the given photon rate, the efficiency of the interferometer (visibility) and detector (quantum efficiency), and the spectral resolution and bandwidth. For our configuration, the single measurement SNR limit of the system is 199. The 5.5-fold discrepancy between the theoretical limit and the experimental value is of purely engineering origin, related to the specific implementation and stability of the optical and mechanical components, and electronics; thus, the performance can be significantly enhanced by refining the implementation.

A similar procedure to outline the typical SNR of classical instruments was performed for a commercial FTIR spectrometer (Bruker LUMOS) with a cryogenically cooled MCT detector; a clear difference in SNR values is observed. For the single-channel measurement, an SNR of 962 was found, determined in the same spectral region between 2900~cm$^{-1}$ and 2400~cm$^{-1}$), whereas for 20 averaged spectra the SNR was 4865.
At first glance, the obtained SNR values for QFTIR and FTIR directly reflect the practical short-term performance. Nevertheless, given the vast difference in the sensing power, i.e., 60~pW (QFTIR, coupled power) versus 1.6~mW (classical FTIR in the same spectral range, measured with a power meter), the feature of the QFTIR method can be better portrayed for the SNR normalized to the IR probing power. 
The SNR normalized to the optical power is $1.5\cdot10^{5}$~mW$^{-1/2}$ for the QFTIR, while for the analyzed classical FTIR it lies at $764$~mW$^{-1/2}$. This means, that for the same optical power for classical and quantum FTIR systems, the SNR will differ by 2-3 orders of magnitude. For this reason, brightness enhancement of SPDC sources is an area of intensive research and development.

It is interesting to note that the shot-noise-limited performance of a classical mid-IR FTIR system (assuming a typical 90\% detector quantum efficiency~\cite{Rogalski2010} and 1.6~mW power within the same band), calculated using the same theoretical framework~\cite{Lindner:21}, yields an SNR of $2.1\cdot10^{5}$. This is more than 217 times higher than the experimental value\textemdash a much larger discrepancy than observed for QFTIR. This highlights that the dominant noise sources are not simply the shot noise of the light itself, but thermal and other intrinsic noises; and that achieving the theoretical limit is significantly more challenging.

\begin{figure}[!ht]
\begin{subfigure}[b]{0.99\linewidth}
         \centering
         \includegraphics[width=\textwidth]{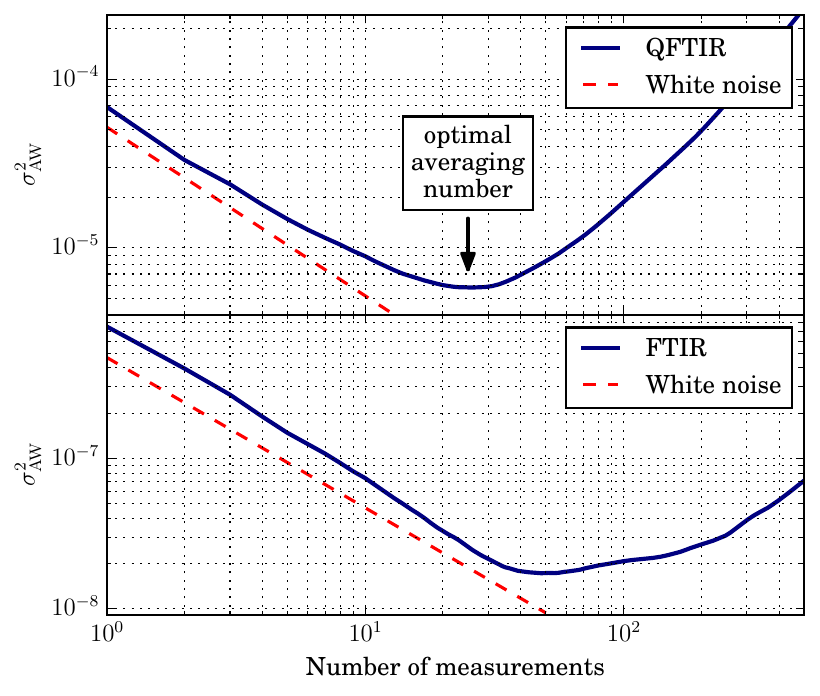}
         \caption{Allan-Werle variance plot for the single-reference mode}
         \label{fig:allan_ratio}
     \end{subfigure}
     \hfill
\begin{subfigure}[b]{0.99\linewidth}
         \centering
         \includegraphics[width=\textwidth]{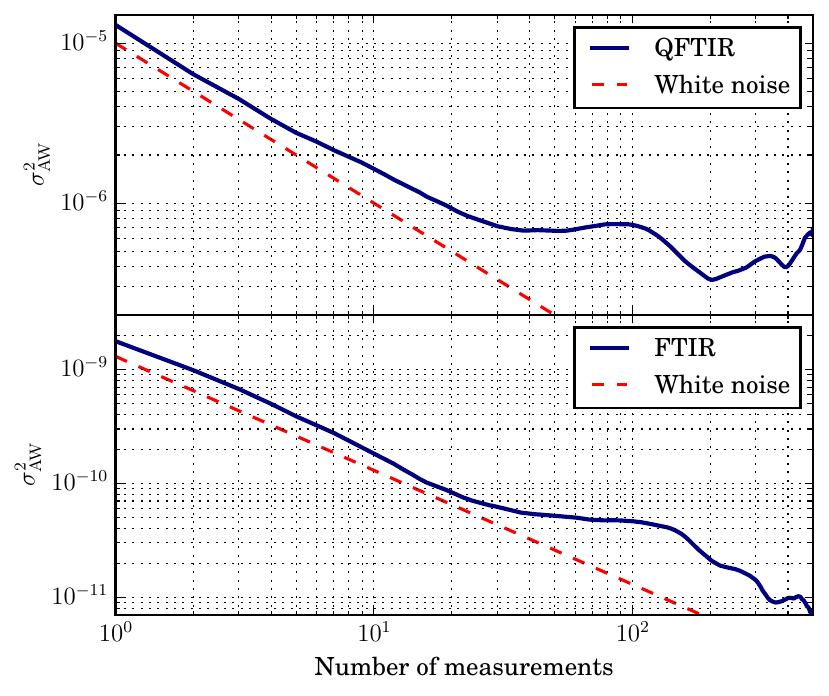}
         \caption{Allan-Werle variance plot for the subsequent referencing mode}
         \label{fig:allan_mean}
     \end{subfigure}
     \hfill
\caption{\label{fig:allan_plots} Long-term performance and stability of QFTIR: overlapping Allan-Werle variance ($\sigma^2_{\mathrm{AW}}$) plots for two spectroscopic measurement modes; the first mode refers to a scenario where the reference spectrum is taken at the beginning of a measurement cycle and stored for absorbance calculations; the second represents a situation where the background (reference) spectrum is recorded sequentially for each consecutive new sample.}
\end{figure}

The long-term stability of the QFTIR spectrometer was quantified by calculating the overlapping Allan-Werle variance~\cite{Werle2011} for two different scenarios. First, a spectroscopic measurement mode, where a reference spectrum is recorded at the beginning of a measurement series, is considered. Therefore, 1000 single-channel transmission spectra (no sample) were recorded, and the integral of the recorded spectral profiles was calculated. 
The overlapping Allan-Werle variance was then computed for the integral values. The result is shown in Fig.~\ref{fig:allan_mean}. In the second case, subsequent reference and sample measurements were taken into account (2000 spectra had to be recorded). The Allan-Werle variance was calculated for the hundred percent lines ($I_n/I_{n+1}$) between two subsequently recorded spectra in the same spectral region as for the prior case.  The result for the second mode is shown in Figure~\ref{fig:allan_ratio}. These findings show, that stable conditions are only present within 25 measurements in the first case (optimal averaging capabilities). The QFTIR system follows the white noise trend limits (shot-noise in the case of QFTIR) but quickly deviates from it. The observed behavior can be attributed to drifts in the system, e.g., temperature drifts causing minor fluctuation in the SPDC emission, or mechanical drifts affecting the single-mode coupling efficiency. In the second case of the subsequent reference and sample measurement modality, more averaging can be justified as the long-term drift of the background is eliminated. However, it must be noted that the Allan variance slightly increases after 60 measurements and has a second minimum at 200 measurements. This behavior can be explained by some instabilities in the experimental setup and long-term drifts in the system.

The long-term performance of QFTIR was also compared to clssical FTIR instruments. For this reason, the same evaluation was applied to spectra recorded with the FTIR (Bruker LUMOS) spectrometer. 2000 measurements were acquired for the Allan-Werle variance evaluation. For the case of consecutive measurements with only a single background measurement made at the beginning of the measurement series, the Allan-Werle plot shows a minimum at 50 measurements. For the case of consecutive background and sample measurements, the LUMOS appeared to be significantly more stable, exhibiting only a slight plato at around 80 measurements and no global minimum in the analyzed range. 

\begin{figure}[hbt]
\centering
\includegraphics[width=0.99\linewidth]{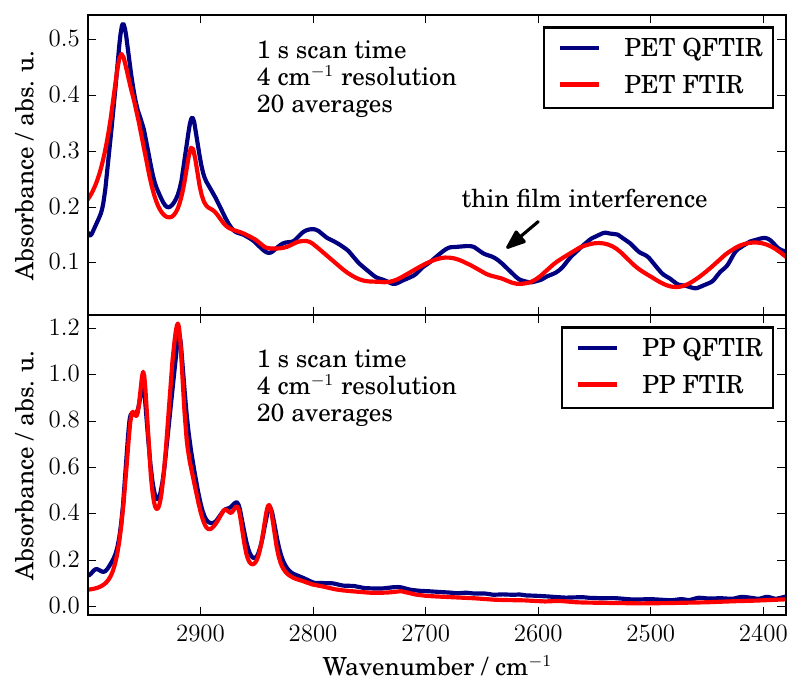}
         \caption{\label{fig:abs_plot} Spectroscopic capabilities of the quantum system (QFTIR) and comparison to the classical instrumentation (FTIR): absorbance spectra (measured in transmission) of polyethylene terephthalate (PET) and polypropylene (PP) thin films; the comparative measurements have been carried out with a standard table-top FTIR spectrometer, the measurement parameters, such as averaging, resolution and speed, were set the same.}
\end{figure}

Besides the short- and long-term analysis, QFTIR measurements with real solid samples were conducted to illustrate the feasibility of the method. Comparative measurements with a classical FTIR were performed for the same samples with the same settings (measurement time, number of averages). 

Thin films of polyethylene terephthalate (PET, 25~\textmu m) and polypropylene (PP, 6~\textmu m) were measured in transmission with the QFTIR setup as well as with a standard tabletop FTIR (Bruker Invenio, uncooled pyroelectric detector). As a reference (background) transmission spectra through air were used. The results, expressed as standard Beer's absorbance, are shown in Figure~\ref{fig:abs_plot}. The comparison measurements of PET show good agreement between the QFTIR and the classical FTIR system (Figure~\ref{fig:abs_plot}). Both measurements were recorded with the same settings and 20 averages per spectrum. Additionally, it was made sure, that a single scan took approximately 1~s. The two measurements agree well with each other since all spectral features can be distinguished; the small variations in the absolute absorbance values can be attributed to different positions in the films that were sampled or slightly different insertion angles. The sinusoidal patterns in the spectra of PET come from thin film interference in the sample, the effect is well-pronounced due to the distinct thickness of the PET films. This information can be used to calculate the film thickness if the group index of the material is known. The PP spectra in Fig.~\ref{fig:abs_plot} agree very well. However small discrepancies at around 3000~cm$^{-1}$ can be attributed to spectral noise close to the emission border. The measurements were performed at aligned polarization in order to minimize polarization effects.



\section{\label{sec:dis}Discussion and outlook}

The recent emergence of quantum sensing techniques as a complement to the well-known gold standard FTIR methods has attracted considerable attention from the classical IR spectroscopy community. In this manuscript, we aimed to objectively evaluate and quantify the current performance of these methods, distinguish their advantages and disadvantages, and relate them to the needs of routine IR spectroscopy. Based on these results, potential perspectives, as well as potential application niches, can be identified.

The QFTIR method enables room-temperature mid-IR spectroscopy, rendering direct IR detectors, sources, and optics unnecessary. This comes with many advantages related to the cost of instrumentation and components, versatility, and ease of construction and operation. 
However, limitations arise due to the specificity of the core component\textemdash the source of non-classical light. Compared to classical FTIR the majority of reported QFTIR spectrometers operate in relatively narrower spectral regions, as their bandwidth is defined by the supported gain profile of the designed SPDC source, i.e., nonlinear crystal. The signal-to-noise ratios of these methods are, as demonstrated in this study, typically 1-2 orders of magnitude lower than their classical counterparts. However, typical QFTIR power levels are also 11-12 orders of magnitude lower. Thus, the amount of spectroscopic information effectively transmitted with each photon is incomparably higher. 
Besides, in the extreme case, QFTIR operates at the shot noise limit, but as we have shown in the study, the drifts of the real implementation soon eliminate this advantage in practice; however, these limitations are purely of engineering origin.
Research groups involved in quantum sensing using undetected photons recognize the above-mentioned limitations and are devoting considerable effort to enhancing the spectral brightness of SPDC emission and broadening the spectral bandwidth.

For instance, to increase the number of generated photons while maintaining broadband emission, pump-enhanced schemes, with a crystal enclosed in a pump-resonant cavity \cite{Lindner2023}, have been reported; the factor of 55 increase in photon number and therefore SNR has been demonstrated. Another alternative route can be the realization of QFTIR in the high-parametric-gain SPDC regime (the pump regimes where signal power starts to exponentially increase with the pump power), providing orders of magnitude larger photon numbers (in the nanowatts to microwatts levels) than standard low-parametric-gain schemes\cite{hashimoto_fourier-transform_2024}. In addition, as demonstrated recently, the realization of high-resolution QFTIR with pulsed sources, previously believed to be unfeasible due to the finite linewidth of the pump pulses and thus broad spectral correlations, can be a simple solution to exploit and harness peak powers of short pulses\cite{Kaur2024} paving the way to high-speed time-resolved spectroscopy. These advances have only recently been demonstrated and more are anticipated.
Regarding spectral coverage, its broadening and extension towards longer wavelengths imply the use of e.g., aperiodic or chirped poling\cite{hashimoto_fourier-transform_2024,Tashima2024} or adoption of other nonlinear materials and systems\cite{Mukai2022, Paterova2022}.

One of the features of the QFTIR technology covered in this study is the low power impinging on the sample, which is in the picowatt range. Therefore, for a given power, QFTIR provides an SNR of about 2 orders of magnitude higher. Thus, no classical FTIR device can compete with QFTIR in areas of study and research where ultra-low sensing energy is a prerequisite. This can be of great interest in applications such as cultural heritage inspection, in-situ biomedical spectroscopy, and imaging (e.g., live cell and tissue analysis), or specific fields where excessive IR radiation can disturb complex systems (e.g., semiconductor industry, analysis of sensitive photoresists, measurements at cryogenic conditions). Hence, QFTIR can be beneficial or the only solution in scenarios where higher power levels could alter or affect the samples. 
It has to be noted, that the QFTIR methods, similar to classical ones, can be advanced to imaging\cite{325251662fcd4d999a1d597579d67827} and polarimetric modalities (due to their intrinsic sensitivity to the polarization changes).
This manuscript also aimed to showcase the current capabilities to a broader range of users and encourage their adoption by applied spectroscopists, which, in turn, would further improve QFTIR methods. 

\section{\label{sec:con} Conclusion}

In this work, we have evaluated and studied the practical performance and applied spectroscopic value of the quantum Fourier transform infrared (QFTIR) spectrometer techniques, covering current possible use cases and extrapolating future prospects. Thus, an experimental QFTIR system in Michelson configuration (based on a 660~nm pump laser and periodically poled potassium titanyl phosphate crystal) has been designed, implemented, and investigated. The experimental setup was compared to standard tabletop instruments \textemdash a Bruker LUMOS and a Bruker Invenio. The spectral noise of the state-of-the-art FTIR system (LUMOS), which uses a nitrogen-cooled MCT detector, is a factor of 27 lower than that of our QFTIR spectrometer which operates with an uncooled silicon detector. However, the power levels of the QFTIR spectrometer were 11 orders of magnitude lower (60~pW versus 1.6~mW), resulting in a power-normalized SNR difference of 2 orders of magnitude ($1.5\cdot10^{5}$~mW$^{-1/2}$ for QFTIR against 764~mW$^{-1/2}$ for classical FTIR).

The long-term stability of QFTIR was also assessed and contrasted with the state-of-the-art system. Allan-Werle variance plots for both systems suggest comparable performance in the measurement mode with a single reference spectrum. The variance minima yield a limit of 25 optimal averages for the QFTIR spectrometer and 50 for the classical instrument.
For the measurement mode of subsequent sampling of spectra and backgrounds, the Allan-Werle plot of the conventional FTIR showed no global minimum for 1000 measurement pairs. Whereas the Allan-Werle plot of the QFTIR system has a global minimum at 200 measurements. The reason for the QFTIR being more prone to drifts and oscillations can be explained by the fact that single-mode fibers were used to couple into the detector. Therefore, slight mechanical drifts in optomechanical components can noticeably deteriorate the signal. Proper engineering and refinement of the QFTIR system design can minimize these effects.

The developed and demonstrated system possesses several constructive features. Particular interest is given to the fact that the combined signal and pump arm (short wave arm) of the nonlinear interferometer is scanned instead of the idler arm. The result is equivalent since the phase differences of the signal and the pump fields are equal to the changes in the idler phase. However, for some technical applications, it is more convenient to have the idler arm fixed \textemdash e.g. when an objective needs to be placed in the idler arm for imaging or mapping modalities.

In the discussion section, the technology's current limitations and strengths with respect to applied spectroscopy are emphasized, along with its advantages, disadvantages, and potential applications. An objective assessment of current performance and future prospects is provided.








\section*{Acknowledgment}

The authors express their gratitude to Christian Rankl for his continued support and his endeavor to initiate this project. Furthermore, the authors want to thank Dominik Kau-Wacht and Bernhard Lendl for discussions of polarization effects in IR spectra. 

This project was co-financed by research subsidies granted by the government of Upper Austria: Quick (Wi-2022-597365/18-Au).

\section*{Declaration}

The authors declared no potential conflicts of interest with respect to the research, authorship, and/or publication of this article.

Data underlying the results presented in this paper are not publicly available at this time but may be obtained from the authors upon reasonable request.

\bibliography{main}

\begin{thebibliography}{32}%
\makeatletter
\providecommand \@ifxundefined [1]{%
 \@ifx{#1\undefined}
}%
\providecommand \@ifnum [1]{%
 \ifnum #1\expandafter \@firstoftwo
 \else \expandafter \@secondoftwo
 \fi
}%
\providecommand \@ifx [1]{%
 \ifx #1\expandafter \@firstoftwo
 \else \expandafter \@secondoftwo
 \fi
}%
\providecommand \natexlab [1]{#1}%
\providecommand \enquote  [1]{``#1''}%
\providecommand \bibnamefont  [1]{#1}%
\providecommand \bibfnamefont [1]{#1}%
\providecommand \citenamefont [1]{#1}%
\providecommand \href@noop [0]{\@secondoftwo}%
\providecommand \href [0]{\begingroup \@sanitize@url \@href}%
\providecommand \@href[1]{\@@startlink{#1}\@@href}%
\providecommand \@@href[1]{\endgroup#1\@@endlink}%
\providecommand \@sanitize@url [0]{\catcode `\\12\catcode `\$12\catcode `\&12\catcode `\#12\catcode `\^12\catcode `\_12\catcode `\%12\relax}%
\providecommand \@@startlink[1]{}%
\providecommand \@@endlink[0]{}%
\providecommand \url  [0]{\begingroup\@sanitize@url \@url }%
\providecommand \@url [1]{\endgroup\@href {#1}{\urlprefix }}%
\providecommand \urlprefix  [0]{URL }%
\providecommand \Eprint [0]{\href }%
\providecommand \doibase [0]{http://dx.doi.org/}%
\providecommand \selectlanguage [0]{\@gobble}%
\providecommand \bibinfo  [0]{\@secondoftwo}%
\providecommand \bibfield  [0]{\@secondoftwo}%
\providecommand \translation [1]{[#1]}%
\providecommand \BibitemOpen [0]{}%
\providecommand \bibitemStop [0]{}%
\providecommand \bibitemNoStop [0]{.\EOS\space}%
\providecommand \EOS [0]{\spacefactor3000\relax}%
\providecommand \BibitemShut  [1]{\csname bibitem#1\endcsname}%
\let\auto@bib@innerbib\@empty
\bibitem [{\citenamefont {Griffiths}\ \emph {et~al.}(2007)\citenamefont {Griffiths}, \citenamefont {De~Haseth},\ and\ \citenamefont {Winefordner}}]{griffiths2007fourier}%
  \BibitemOpen
  \bibfield  {author} {\bibinfo {author} {\bibfnamefont {P.}~\bibnamefont {Griffiths}}, \bibinfo {author} {\bibfnamefont {J.}~\bibnamefont {De~Haseth}}, \ and\ \bibinfo {author} {\bibfnamefont {J.}~\bibnamefont {Winefordner}},\ }\href {https://books.google.at/books?id=ZecrNiUkHToC} {\emph {\bibinfo {title} {Fourier Transform Infrared Spectrometry}}},\ Chemical Analysis: A Series of Monographs on Analytical Chemistry and Its Applications\ (\bibinfo  {publisher} {Wiley},\ \bibinfo {year} {2007})\BibitemShut {NoStop}%
\bibitem [{\citenamefont {Chalmers}\ and\ \citenamefont {Griffiths}(2002)}]{chalmers_handbook_2001}%
  \BibitemOpen
  \bibinfo {editor} {\bibfnamefont {J.~M.}\ \bibnamefont {Chalmers}}\ and\ \bibinfo {editor} {\bibfnamefont {P.~R.}\ \bibnamefont {Griffiths}},\ eds.,\ \href {\doibase 10.1002/0470027320} {\emph {\bibinfo {title} {Handbook of Vibrational Spectroscopy}}}\ (\bibinfo  {publisher} {John Wiley \& Sons, Ltd},\ \bibinfo {address} {Chichester, UK},\ \bibinfo {year} {2002})\BibitemShut {NoStop}%
\bibitem [{\citenamefont {Zorin}\ \emph {et~al.}(2022)\citenamefont {Zorin}, \citenamefont {Gattinger}, \citenamefont {Ebner},\ and\ \citenamefont {Brandstetter}}]{Zorin:22}%
  \BibitemOpen
  \bibfield  {author} {\bibinfo {author} {\bibfnamefont {I.}~\bibnamefont {Zorin}}, \bibinfo {author} {\bibfnamefont {P.}~\bibnamefont {Gattinger}}, \bibinfo {author} {\bibfnamefont {A.}~\bibnamefont {Ebner}}, \ and\ \bibinfo {author} {\bibfnamefont {M.}~\bibnamefont {Brandstetter}},\ }\href {\doibase 10.1364/OE.447269} {\bibfield  {journal} {\bibinfo  {journal} {Opt. Express}\ }\textbf {\bibinfo {volume} {30}},\ \bibinfo {pages} {5222} (\bibinfo {year} {2022})}\BibitemShut {NoStop}%
\bibitem [{\citenamefont {Krebbers}\ \emph {et~al.}(2024)\citenamefont {Krebbers}, \citenamefont {van Kempen}, \citenamefont {Harren}, \citenamefont {Vasilyev}, \citenamefont {Peterse}, \citenamefont {L\"{u}cker}, \citenamefont {Khodabakhsh},\ and\ \citenamefont {Cristescu}}]{Krebbers:24}%
  \BibitemOpen
  \bibfield  {author} {\bibinfo {author} {\bibfnamefont {R.}~\bibnamefont {Krebbers}}, \bibinfo {author} {\bibfnamefont {K.}~\bibnamefont {van Kempen}}, \bibinfo {author} {\bibfnamefont {F.~J.~M.}\ \bibnamefont {Harren}}, \bibinfo {author} {\bibfnamefont {S.}~\bibnamefont {Vasilyev}}, \bibinfo {author} {\bibfnamefont {I.~F.}\ \bibnamefont {Peterse}}, \bibinfo {author} {\bibfnamefont {S.}~\bibnamefont {L\"{u}cker}}, \bibinfo {author} {\bibfnamefont {A.}~\bibnamefont {Khodabakhsh}}, \ and\ \bibinfo {author} {\bibfnamefont {S.~M.}\ \bibnamefont {Cristescu}},\ }\href {\doibase 10.1364/OE.515914} {\bibfield  {journal} {\bibinfo  {journal} {Opt. Express}\ }\textbf {\bibinfo {volume} {32}},\ \bibinfo {pages} {14506} (\bibinfo {year} {2024})}\BibitemShut {NoStop}%
\bibitem [{\citenamefont {Rogalski}(2010)}]{Rogalski2010}%
  \BibitemOpen
  \bibfield  {author} {\bibinfo {author} {\bibfnamefont {A.}~\bibnamefont {Rogalski}},\ }\href {\doibase 10.1201/b10319} {\emph {\bibinfo {title} {{Infrared Detectors}}}}\ (\bibinfo  {publisher} {CRC Press},\ \bibinfo {year} {2010})\ p.\ \bibinfo {pages} {898}\BibitemShut {NoStop}%
\bibitem [{\citenamefont {Rogalski}\ \emph {et~al.}(2021)\citenamefont {Rogalski}, \citenamefont {Martyniuk}, \citenamefont {Kopytko},\ and\ \citenamefont {Hu}}]{Rogalski2021}%
  \BibitemOpen
  \bibfield  {author} {\bibinfo {author} {\bibfnamefont {A.}~\bibnamefont {Rogalski}}, \bibinfo {author} {\bibfnamefont {P.}~\bibnamefont {Martyniuk}}, \bibinfo {author} {\bibfnamefont {M.}~\bibnamefont {Kopytko}}, \ and\ \bibinfo {author} {\bibfnamefont {W.}~\bibnamefont {Hu}},\ }\href {\doibase 10.3390/app11020501} {\bibfield  {journal} {\bibinfo  {journal} {Applied Sciences}\ }\textbf {\bibinfo {volume} {11}},\ \bibinfo {pages} {501} (\bibinfo {year} {2021})}\BibitemShut {NoStop}%
\bibitem [{\citenamefont {Konstantatos}(2018)}]{Konstantatos2018}%
  \BibitemOpen
  \bibfield  {author} {\bibinfo {author} {\bibfnamefont {G.}~\bibnamefont {Konstantatos}},\ }\href {\doibase 10.1038/s41467-018-07643-7} {\bibfield  {journal} {\bibinfo  {journal} {Nature Communications}\ }\textbf {\bibinfo {volume} {9}},\ \bibinfo {pages} {5266} (\bibinfo {year} {2018})}\BibitemShut {NoStop}%
\bibitem [{\citenamefont {Chekhova}\ and\ \citenamefont {Ou}(2016)}]{Chekhova2016}%
  \BibitemOpen
  \bibfield  {author} {\bibinfo {author} {\bibfnamefont {M.~V.}\ \bibnamefont {Chekhova}}\ and\ \bibinfo {author} {\bibfnamefont {Z.~Y.}\ \bibnamefont {Ou}},\ }\href {\doibase 10.1364/AOP.8.000104} {\bibfield  {journal} {\bibinfo  {journal} {Advances in Optics and Photonics}\ }\textbf {\bibinfo {volume} {8}},\ \bibinfo {pages} {104} (\bibinfo {year} {2016})}\BibitemShut {NoStop}%
\bibitem [{\citenamefont {Zou}\ \emph {et~al.}(1991)\citenamefont {Zou}, \citenamefont {Wang},\ and\ \citenamefont {Mandel}}]{Zou1991}%
  \BibitemOpen
  \bibfield  {author} {\bibinfo {author} {\bibfnamefont {X.~Y.}\ \bibnamefont {Zou}}, \bibinfo {author} {\bibfnamefont {L.~J.}\ \bibnamefont {Wang}}, \ and\ \bibinfo {author} {\bibfnamefont {L.}~\bibnamefont {Mandel}},\ }\href {\doibase 10.1103/PhysRevLett.67.318} {\bibfield  {journal} {\bibinfo  {journal} {Physical Review Letters}\ }\textbf {\bibinfo {volume} {67}},\ \bibinfo {pages} {318} (\bibinfo {year} {1991})}\BibitemShut {NoStop}%
\bibitem [{\citenamefont {Mandel}(1991)}]{Mandel1991}%
  \BibitemOpen
  \bibfield  {author} {\bibinfo {author} {\bibfnamefont {L.}~\bibnamefont {Mandel}},\ }\href {\doibase 10.1364/OL.16.001882} {\bibfield  {journal} {\bibinfo  {journal} {Optics Letters}\ }\textbf {\bibinfo {volume} {16}},\ \bibinfo {pages} {1882} (\bibinfo {year} {1991})}\BibitemShut {NoStop}%
\bibitem [{\citenamefont {Lemos}\ \emph {et~al.}(2014)\citenamefont {Lemos}, \citenamefont {Borish}, \citenamefont {Cole}, \citenamefont {Ramelow}, \citenamefont {Lapkiewicz},\ and\ \citenamefont {Zeilinger}}]{Lemos2014}%
  \BibitemOpen
  \bibfield  {author} {\bibinfo {author} {\bibfnamefont {G.~B.}\ \bibnamefont {Lemos}}, \bibinfo {author} {\bibfnamefont {V.}~\bibnamefont {Borish}}, \bibinfo {author} {\bibfnamefont {G.~D.}\ \bibnamefont {Cole}}, \bibinfo {author} {\bibfnamefont {S.}~\bibnamefont {Ramelow}}, \bibinfo {author} {\bibfnamefont {R.}~\bibnamefont {Lapkiewicz}}, \ and\ \bibinfo {author} {\bibfnamefont {A.}~\bibnamefont {Zeilinger}},\ }\href {\doibase 10.1038/nature13586} {\bibfield  {journal} {\bibinfo  {journal} {Nature 2014 512:7515}\ }\textbf {\bibinfo {volume} {512}},\ \bibinfo {pages} {409} (\bibinfo {year} {2014})}\BibitemShut {NoStop}%
\bibitem [{\citenamefont {{Barreto Lemos}}\ \emph {et~al.}(2022)\citenamefont {{Barreto Lemos}}, \citenamefont {Lahiri}, \citenamefont {Ramelow}, \citenamefont {Lapkiewicz},\ and\ \citenamefont {Plick}}]{BarretoLemos2022}%
  \BibitemOpen
  \bibfield  {author} {\bibinfo {author} {\bibfnamefont {G.}~\bibnamefont {{Barreto Lemos}}}, \bibinfo {author} {\bibfnamefont {M.}~\bibnamefont {Lahiri}}, \bibinfo {author} {\bibfnamefont {S.}~\bibnamefont {Ramelow}}, \bibinfo {author} {\bibfnamefont {R.}~\bibnamefont {Lapkiewicz}}, \ and\ \bibinfo {author} {\bibfnamefont {W.~N.}\ \bibnamefont {Plick}},\ }\href {\doibase 10.1364/JOSAB.456778} {\bibfield  {journal} {\bibinfo  {journal} {Journal of the Optical Society of America B}\ }\textbf {\bibinfo {volume} {39}},\ \bibinfo {pages} {2200} (\bibinfo {year} {2022})}\BibitemShut {NoStop}%
\bibitem [{\citenamefont {Kalashnikov}\ \emph {et~al.}(2016)\citenamefont {Kalashnikov}, \citenamefont {Paterova}, \citenamefont {Kulik},\ and\ \citenamefont {Krivitsky}}]{Kalashnikov2016}%
  \BibitemOpen
  \bibfield  {author} {\bibinfo {author} {\bibfnamefont {D.~A.}\ \bibnamefont {Kalashnikov}}, \bibinfo {author} {\bibfnamefont {A.~V.}\ \bibnamefont {Paterova}}, \bibinfo {author} {\bibfnamefont {S.~P.}\ \bibnamefont {Kulik}}, \ and\ \bibinfo {author} {\bibfnamefont {L.~A.}\ \bibnamefont {Krivitsky}},\ }\href {\doibase 10.1038/nphoton.2015.252} {\bibfield  {journal} {\bibinfo  {journal} {Nature Photonics}\ }\textbf {\bibinfo {volume} {10}},\ \bibinfo {pages} {98} (\bibinfo {year} {2016})}\BibitemShut {NoStop}%
\bibitem [{\citenamefont {Paterova}\ \emph {et~al.}(2022)\citenamefont {Paterova}, \citenamefont {Toa}, \citenamefont {Yang},\ and\ \citenamefont {Krivitsky}}]{Paterova2022}%
  \BibitemOpen
  \bibfield  {author} {\bibinfo {author} {\bibfnamefont {A.~V.}\ \bibnamefont {Paterova}}, \bibinfo {author} {\bibfnamefont {Z.~S.~D.}\ \bibnamefont {Toa}}, \bibinfo {author} {\bibfnamefont {H.}~\bibnamefont {Yang}}, \ and\ \bibinfo {author} {\bibfnamefont {L.~A.}\ \bibnamefont {Krivitsky}},\ }\href {\doibase 10.1021/acsphotonics.2c00464} {\bibfield  {journal} {\bibinfo  {journal} {ACS Photonics}\ }\textbf {\bibinfo {volume} {9}},\ \bibinfo {pages} {2151} (\bibinfo {year} {2022})}\BibitemShut {NoStop}%
\bibitem [{\citenamefont {Kaufmann}\ \emph {et~al.}(2022)\citenamefont {Kaufmann}, \citenamefont {Chrzanowski}, \citenamefont {Vanselow},\ and\ \citenamefont {Ramelow}}]{Kaufmann2022}%
  \BibitemOpen
  \bibfield  {author} {\bibinfo {author} {\bibfnamefont {P.}~\bibnamefont {Kaufmann}}, \bibinfo {author} {\bibfnamefont {H.~M.}\ \bibnamefont {Chrzanowski}}, \bibinfo {author} {\bibfnamefont {A.}~\bibnamefont {Vanselow}}, \ and\ \bibinfo {author} {\bibfnamefont {S.}~\bibnamefont {Ramelow}},\ }\href {\doibase 10.1364/OE.442411} {\bibfield  {journal} {\bibinfo  {journal} {Optics Express}\ }\textbf {\bibinfo {volume} {30}},\ \bibinfo {pages} {5926} (\bibinfo {year} {2022})}\BibitemShut {NoStop}%
\bibitem [{\citenamefont {Cardoso}\ \emph {et~al.}(2024)\citenamefont {Cardoso}, \citenamefont {Dong}, \citenamefont {Zhou}, \citenamefont {Joshi},\ and\ \citenamefont {Rarity}}]{Cardoso2024}%
  \BibitemOpen
  \bibfield  {author} {\bibinfo {author} {\bibfnamefont {A.~C.}\ \bibnamefont {Cardoso}}, \bibinfo {author} {\bibfnamefont {J.}~\bibnamefont {Dong}}, \bibinfo {author} {\bibfnamefont {H.}~\bibnamefont {Zhou}}, \bibinfo {author} {\bibfnamefont {S.~K.}\ \bibnamefont {Joshi}}, \ and\ \bibinfo {author} {\bibfnamefont {J.~G.}\ \bibnamefont {Rarity}},\ }\href {\doibase 10.1364/OPTCON.524280} {\bibfield  {journal} {\bibinfo  {journal} {Optics Continuum}\ }\textbf {\bibinfo {volume} {3}},\ \bibinfo {pages} {823} (\bibinfo {year} {2024})}\BibitemShut {NoStop}%
\bibitem [{\citenamefont {Vall{\'{e}}s}\ \emph {et~al.}(2018)\citenamefont {Vall{\'{e}}s}, \citenamefont {Jim{\'{e}}nez}, \citenamefont {Salazar-Serrano},\ and\ \citenamefont {Torres}}]{Valles2018}%
  \BibitemOpen
  \bibfield  {author} {\bibinfo {author} {\bibfnamefont {A.}~\bibnamefont {Vall{\'{e}}s}}, \bibinfo {author} {\bibfnamefont {G.}~\bibnamefont {Jim{\'{e}}nez}}, \bibinfo {author} {\bibfnamefont {L.~J.}\ \bibnamefont {Salazar-Serrano}}, \ and\ \bibinfo {author} {\bibfnamefont {J.~P.}\ \bibnamefont {Torres}},\ }\href {\doibase 10.1103/PhysRevA.97.023824} {\bibfield  {journal} {\bibinfo  {journal} {Physical Review A}\ }\textbf {\bibinfo {volume} {97}},\ \bibinfo {pages} {023824} (\bibinfo {year} {2018})}\BibitemShut {NoStop}%
\bibitem [{\citenamefont {Paterova}\ \emph {et~al.}(2018)\citenamefont {Paterova}, \citenamefont {Yang}, \citenamefont {An}, \citenamefont {Kalashnikov},\ and\ \citenamefont {Krivitsky}}]{Paterova2018}%
  \BibitemOpen
  \bibfield  {author} {\bibinfo {author} {\bibfnamefont {A.~V.}\ \bibnamefont {Paterova}}, \bibinfo {author} {\bibfnamefont {H.}~\bibnamefont {Yang}}, \bibinfo {author} {\bibfnamefont {C.}~\bibnamefont {An}}, \bibinfo {author} {\bibfnamefont {D.~A.}\ \bibnamefont {Kalashnikov}}, \ and\ \bibinfo {author} {\bibfnamefont {L.~A.}\ \bibnamefont {Krivitsky}},\ }\href {\doibase 10.1088/2058-9565/aab567} {\bibfield  {journal} {\bibinfo  {journal} {Quantum Science and Technology}\ }\textbf {\bibinfo {volume} {3}},\ \bibinfo {pages} {025008} (\bibinfo {year} {2018})}\BibitemShut {NoStop}%
\bibitem [{\citenamefont {Vanselow}\ \emph {et~al.}(2020)\citenamefont {Vanselow}, \citenamefont {Kaufmann}, \citenamefont {Zorin}, \citenamefont {Heise}, \citenamefont {Chrzanowski},\ and\ \citenamefont {Ramelow}}]{Vanselow2020}%
  \BibitemOpen
  \bibfield  {author} {\bibinfo {author} {\bibfnamefont {A.}~\bibnamefont {Vanselow}}, \bibinfo {author} {\bibfnamefont {P.}~\bibnamefont {Kaufmann}}, \bibinfo {author} {\bibfnamefont {I.}~\bibnamefont {Zorin}}, \bibinfo {author} {\bibfnamefont {B.}~\bibnamefont {Heise}}, \bibinfo {author} {\bibfnamefont {H.~M.}\ \bibnamefont {Chrzanowski}}, \ and\ \bibinfo {author} {\bibfnamefont {S.}~\bibnamefont {Ramelow}},\ }\href {\doibase 10.1364/OPTICA.400128} {\bibfield  {journal} {\bibinfo  {journal} {Optica, Vol. 7, Issue 12, pp. 1729-1736}\ }\textbf {\bibinfo {volume} {7}},\ \bibinfo {pages} {1729} (\bibinfo {year} {2020})}\BibitemShut {NoStop}%
\bibitem [{\citenamefont {Lindner}\ \emph {et~al.}(2020)\citenamefont {Lindner}, \citenamefont {Wolf}, \citenamefont {Kiessling},\ and\ \citenamefont {K{\"{u}}hnemann}}]{Lindner2020}%
  \BibitemOpen
  \bibfield  {author} {\bibinfo {author} {\bibfnamefont {C.}~\bibnamefont {Lindner}}, \bibinfo {author} {\bibfnamefont {S.}~\bibnamefont {Wolf}}, \bibinfo {author} {\bibfnamefont {J.}~\bibnamefont {Kiessling}}, \ and\ \bibinfo {author} {\bibfnamefont {F.}~\bibnamefont {K{\"{u}}hnemann}},\ }\href {\doibase 10.1364/OE.382351} {\bibfield  {journal} {\bibinfo  {journal} {Optics Express}\ }\textbf {\bibinfo {volume} {28}},\ \bibinfo {pages} {4426} (\bibinfo {year} {2020})}\BibitemShut {NoStop}%
\bibitem [{\citenamefont {Lindner}\ \emph {et~al.}(2021)\citenamefont {Lindner}, \citenamefont {Kunz}, \citenamefont {Herr}, \citenamefont {Wolf}, \citenamefont {Kie{\ss}ling},\ and\ \citenamefont {K{\"{u}}hnemann}}]{Lindner:21}%
  \BibitemOpen
  \bibfield  {author} {\bibinfo {author} {\bibfnamefont {C.}~\bibnamefont {Lindner}}, \bibinfo {author} {\bibfnamefont {J.}~\bibnamefont {Kunz}}, \bibinfo {author} {\bibfnamefont {S.~J.}\ \bibnamefont {Herr}}, \bibinfo {author} {\bibfnamefont {S.}~\bibnamefont {Wolf}}, \bibinfo {author} {\bibfnamefont {J.}~\bibnamefont {Kie{\ss}ling}}, \ and\ \bibinfo {author} {\bibfnamefont {F.}~\bibnamefont {K{\"{u}}hnemann}},\ }\href {\doibase 10.1364/OE.415365} {\bibfield  {journal} {\bibinfo  {journal} {Opt. Express}\ }\textbf {\bibinfo {volume} {29}},\ \bibinfo {pages} {4035} (\bibinfo {year} {2021})}\BibitemShut {NoStop}%
\bibitem [{\citenamefont {Lindner}\ \emph {et~al.}(2022)\citenamefont {Lindner}, \citenamefont {Kunz}, \citenamefont {Herr}, \citenamefont {Kiessling}, \citenamefont {Wolf},\ and\ \citenamefont {K{\"{u}}hnemann}}]{Lindner2022}%
  \BibitemOpen
  \bibfield  {author} {\bibinfo {author} {\bibfnamefont {C.}~\bibnamefont {Lindner}}, \bibinfo {author} {\bibfnamefont {J.}~\bibnamefont {Kunz}}, \bibinfo {author} {\bibfnamefont {S.~J.}\ \bibnamefont {Herr}}, \bibinfo {author} {\bibfnamefont {J.}~\bibnamefont {Kiessling}}, \bibinfo {author} {\bibfnamefont {S.}~\bibnamefont {Wolf}}, \ and\ \bibinfo {author} {\bibfnamefont {F.}~\bibnamefont {K{\"{u}}hnemann}},\ }\href {\doibase 10.1364/OPTCON.449219} {\bibfield  {journal} {\bibinfo  {journal} {Optics Continuum}\ }\textbf {\bibinfo {volume} {1}},\ \bibinfo {pages} {189} (\bibinfo {year} {2022})}\BibitemShut {NoStop}%
\bibitem [{\citenamefont {Mukai}\ \emph {et~al.}(2022)\citenamefont {Mukai}, \citenamefont {Okamoto},\ and\ \citenamefont {Takeuchi}}]{Mukai2022}%
  \BibitemOpen
  \bibfield  {author} {\bibinfo {author} {\bibfnamefont {Y.}~\bibnamefont {Mukai}}, \bibinfo {author} {\bibfnamefont {R.}~\bibnamefont {Okamoto}}, \ and\ \bibinfo {author} {\bibfnamefont {S.}~\bibnamefont {Takeuchi}},\ }\href {\doibase 10.1364/OE.455718} {\bibfield  {journal} {\bibinfo  {journal} {Optics Express}\ }\textbf {\bibinfo {volume} {30}},\ \bibinfo {pages} {22624} (\bibinfo {year} {2022})}\BibitemShut {NoStop}%
\bibitem [{\citenamefont {Tashima}\ \emph {et~al.}(2024)\citenamefont {Tashima}, \citenamefont {Mukai}, \citenamefont {Arahata}, \citenamefont {Oda}, \citenamefont {Hisamitsu}, \citenamefont {Tokuda}, \citenamefont {Okamoto},\ and\ \citenamefont {Takeuchi}}]{Tashima2024}%
  \BibitemOpen
  \bibfield  {author} {\bibinfo {author} {\bibfnamefont {T.}~\bibnamefont {Tashima}}, \bibinfo {author} {\bibfnamefont {Y.}~\bibnamefont {Mukai}}, \bibinfo {author} {\bibfnamefont {M.}~\bibnamefont {Arahata}}, \bibinfo {author} {\bibfnamefont {N.}~\bibnamefont {Oda}}, \bibinfo {author} {\bibfnamefont {M.}~\bibnamefont {Hisamitsu}}, \bibinfo {author} {\bibfnamefont {K.}~\bibnamefont {Tokuda}}, \bibinfo {author} {\bibfnamefont {R.}~\bibnamefont {Okamoto}}, \ and\ \bibinfo {author} {\bibfnamefont {S.}~\bibnamefont {Takeuchi}},\ }\href {\doibase 10.1364/OPTICA.504450} {\bibfield  {journal} {\bibinfo  {journal} {Optica}\ }\textbf {\bibinfo {volume} {11}},\ \bibinfo {pages} {81} (\bibinfo {year} {2024})}\BibitemShut {NoStop}%
\bibitem [{\citenamefont {Vanselow}\ \emph {et~al.}(2019)\citenamefont {Vanselow}, \citenamefont {Kaufmann}, \citenamefont {Chrzanowski},\ and\ \citenamefont {Ramelow}}]{Vanselow2019}%
  \BibitemOpen
  \bibfield  {author} {\bibinfo {author} {\bibfnamefont {A.}~\bibnamefont {Vanselow}}, \bibinfo {author} {\bibfnamefont {P.}~\bibnamefont {Kaufmann}}, \bibinfo {author} {\bibfnamefont {H.~M.}\ \bibnamefont {Chrzanowski}}, \ and\ \bibinfo {author} {\bibfnamefont {S.}~\bibnamefont {Ramelow}},\ }\href {\doibase 10.1364/OL.44.004638} {\bibfield  {journal} {\bibinfo  {journal} {Optics Letters}\ }\textbf {\bibinfo {volume} {44}},\ \bibinfo {pages} {4638} (\bibinfo {year} {2019})}\BibitemShut {NoStop}%
\bibitem [{\citenamefont {Placke}\ \emph {et~al.}(2023)\citenamefont {Placke}, \citenamefont {Lindner}, \citenamefont {Kviatkovsky}, \citenamefont {Chrzanowski}, \citenamefont {K{\"{u}}hnemann},\ and\ \citenamefont {Ramelow}}]{Placke2023}%
  \BibitemOpen
  \bibfield  {author} {\bibinfo {author} {\bibfnamefont {M.}~\bibnamefont {Placke}}, \bibinfo {author} {\bibfnamefont {C.}~\bibnamefont {Lindner}}, \bibinfo {author} {\bibfnamefont {I.}~\bibnamefont {Kviatkovsky}}, \bibinfo {author} {\bibfnamefont {H.~M.}\ \bibnamefont {Chrzanowski}}, \bibinfo {author} {\bibfnamefont {F.}~\bibnamefont {K{\"{u}}hnemann}}, \ and\ \bibinfo {author} {\bibfnamefont {S.}~\bibnamefont {Ramelow}},\ }in\ \href {\doibase 10.1364/CLEO_AT.2023.AM2N.4} {\emph {\bibinfo {booktitle} {CLEO 2023}}}\ (\bibinfo  {publisher} {Optica Publishing Group},\ \bibinfo {address} {Washington, D.C.},\ \bibinfo {year} {2023})\ p.\ \bibinfo {pages} {AM2N.4}\BibitemShut {NoStop}%
\bibitem [{\citenamefont {Saptari}(2003)}]{Saptari2003}%
  \BibitemOpen
  \bibfield  {author} {\bibinfo {author} {\bibfnamefont {V.}~\bibnamefont {Saptari}},\ }\href {\doibase 10.1117/3.523499} {\emph {\bibinfo {title} {{Fourier-Transform Spectroscopy Instrumentation Engineering}}}}\ (\bibinfo  {publisher} {SPIE},\ \bibinfo {year} {2003})\ p.\ \bibinfo {pages} {136}\BibitemShut {NoStop}%
\bibitem [{\citenamefont {Werle}(2011)}]{Werle2011}%
  \BibitemOpen
  \bibfield  {author} {\bibinfo {author} {\bibfnamefont {P.}~\bibnamefont {Werle}},\ }\href {\doibase 10.1007/s00340-010-4165-9} {\bibfield  {journal} {\bibinfo  {journal} {Applied Physics B}\ }\textbf {\bibinfo {volume} {102}},\ \bibinfo {pages} {313} (\bibinfo {year} {2011})}\BibitemShut {NoStop}%
\bibitem [{\citenamefont {Lindner}\ \emph {et~al.}(2023)\citenamefont {Lindner}, \citenamefont {Kunz}, \citenamefont {Herr}, \citenamefont {Kie{\ss}ling}, \citenamefont {Wolf},\ and\ \citenamefont {K{\"{u}}hnemann}}]{Lindner2023}%
  \BibitemOpen
  \bibfield  {author} {\bibinfo {author} {\bibfnamefont {C.}~\bibnamefont {Lindner}}, \bibinfo {author} {\bibfnamefont {J.}~\bibnamefont {Kunz}}, \bibinfo {author} {\bibfnamefont {S.~J.}\ \bibnamefont {Herr}}, \bibinfo {author} {\bibfnamefont {J.}~\bibnamefont {Kie{\ss}ling}}, \bibinfo {author} {\bibfnamefont {S.}~\bibnamefont {Wolf}}, \ and\ \bibinfo {author} {\bibfnamefont {F.}~\bibnamefont {K{\"{u}}hnemann}},\ }\href {\doibase 10.1063/5.0146025} {\bibfield  {journal} {\bibinfo  {journal} {APL Photonics}\ }\textbf {\bibinfo {volume} {8}} (\bibinfo {year} {2023}),\ 10.1063/5.0146025}\BibitemShut {NoStop}%
\bibitem [{\citenamefont {Hashimoto}\ \emph {et~al.}(2024)\citenamefont {Hashimoto}, \citenamefont {Horoshko}, \citenamefont {Kolobov}, \citenamefont {Michael}, \citenamefont {Gefen},\ and\ \citenamefont {Chekhova}}]{hashimoto_fourier-transform_2024}%
  \BibitemOpen
  \bibfield  {author} {\bibinfo {author} {\bibfnamefont {K.}~\bibnamefont {Hashimoto}}, \bibinfo {author} {\bibfnamefont {D.~B.}\ \bibnamefont {Horoshko}}, \bibinfo {author} {\bibfnamefont {M.~I.}\ \bibnamefont {Kolobov}}, \bibinfo {author} {\bibfnamefont {Y.}~\bibnamefont {Michael}}, \bibinfo {author} {\bibfnamefont {Z.}~\bibnamefont {Gefen}}, \ and\ \bibinfo {author} {\bibfnamefont {M.~V.}\ \bibnamefont {Chekhova}},\ }\href {\doibase 10.1038/s42005-024-01717-3} {\bibfield  {journal} {\bibinfo  {journal} {Communications Physics}\ }\textbf {\bibinfo {volume} {7}},\ \bibinfo {pages} {217} (\bibinfo {year} {2024})}\BibitemShut {NoStop}%
\bibitem [{\citenamefont {Kaur}\ \emph {et~al.}(2024)\citenamefont {Kaur}, \citenamefont {Mukai}, \citenamefont {Okamoto},\ and\ \citenamefont {Takeuchi}}]{Kaur2024}%
  \BibitemOpen
  \bibfield  {author} {\bibinfo {author} {\bibfnamefont {J.}~\bibnamefont {Kaur}}, \bibinfo {author} {\bibfnamefont {Y.}~\bibnamefont {Mukai}}, \bibinfo {author} {\bibfnamefont {R.}~\bibnamefont {Okamoto}}, \ and\ \bibinfo {author} {\bibfnamefont {S.}~\bibnamefont {Takeuchi}},\ }\href {\doibase 10.1103/PhysRevApplied.22.044015} {\bibfield  {journal} {\bibinfo  {journal} {Physical Review Applied}\ }\textbf {\bibinfo {volume} {22}},\ \bibinfo {pages} {044015} (\bibinfo {year} {2024})}\BibitemShut {NoStop}%
\bibitem [{\citenamefont {Defienne}\ \emph {et~al.}(2024)\citenamefont {Defienne}, \citenamefont {Bowen}, \citenamefont {Chekhova}, \citenamefont {Lemos}, \citenamefont {Oron}, \citenamefont {Ramelow}, \citenamefont {Treps},\ and\ \citenamefont {Faccio}}]{325251662fcd4d999a1d597579d67827}%
  \BibitemOpen
  \bibfield  {author} {\bibinfo {author} {\bibfnamefont {H.}~\bibnamefont {Defienne}}, \bibinfo {author} {\bibfnamefont {W.}~\bibnamefont {Bowen}}, \bibinfo {author} {\bibfnamefont {M.}~\bibnamefont {Chekhova}}, \bibinfo {author} {\bibfnamefont {G.}~\bibnamefont {Lemos}}, \bibinfo {author} {\bibfnamefont {D.}~\bibnamefont {Oron}}, \bibinfo {author} {\bibfnamefont {S.}~\bibnamefont {Ramelow}}, \bibinfo {author} {\bibfnamefont {N.}~\bibnamefont {Treps}}, \ and\ \bibinfo {author} {\bibfnamefont {D.}~\bibnamefont {Faccio}},\ }\href {\doibase 10.1038/s41566-024-01516-w} {\bibfield  {journal} {\bibinfo  {journal} {Nature Photonics}\ }\textbf {\bibinfo {volume} {18}},\ \bibinfo {pages} {1024} (\bibinfo {year} {2024})},\ \bibinfo {note} {publisher Copyright: {\textcopyright} Springer Nature Limited 2024.}\BibitemShut {Stop}%
\end{thebibliography}%
\end{document}